\def\half{\frac{1}{2}}

\def\bey{\begin{eqnarray}}
\def\eey{\end{eqnarray}}
\def\be{\begin{equation}}
\def\ee{\end{equation}}
\def\ba{\begin{array}}
\def\ea{\end{array}}
\def\gm{\gamma}

\def\ld{\lambda}

\def\af{\alpha}
\def\sg{\sigma}
\def\Sg{\Sigma}

\def\om{\omega}
\def\r{\rho}

\def\bt{\beta}

\def\dt{\delta}
\def\Dt{\Delta}

\def\pp{\partial}
\def\pp{\partial}

\def\nnb{\nonumber}
\documentstyle[11pt,epsfig]{article}
\title{Chiral condensate in nuclear matter with vacuum corrections}
\author{Wei-Zhou Jiang and  Bao-An Li \\
\it  Department of Physics, Texas A\&M University-Commerce, Commerce,
TX 75429, USA}
\date{}
\bigskip
\begin{document}
\maketitle \baselineskip 18.6pt
\begin{abstract}
\baselineskip14pt Within the relativistic Hartree approach using a
Lagrangian with density-dependent parameters according to the
Brown-Rho scaling law, it is found that the vacuum corrections from
the nucleon Dirac sea soften the equation of state and favor the
chiral symmetry restoration at high densities.

\thanks Keywords: Chiral condensate,
Relativistic Hartree approach, nuclear matter; PACS number:24.85.+p,
11.30.Rd, 21.60.Jz, 21.65.+f
\end{abstract}

\section{Introduction}
As hadrons can be viewed as excitations of the QCD vacuum, the vacuum
structure that features varieties of quark and gluon condensates
plays a crucial role in the hadron
dynamics~\cite{vo97,ko97,ca99,dr01,br02}. The chiral or quark
condensate $<\bar{q}q>$ serves as the order parameter for the
spontaneous chiral symmetry breaking and reflects the nature of the
QCD vacuum. For a chirally invariant vacuum, the chiral condensate
vanishes. The non-vanishing chiral condensate plays a crucial role in
generating the dynamical chiral symmetry breaking with the mass
acquisition of constituent quarks.  In free space,
$<\bar{q}q>_0=-(225\pm25~ \rm{MeV})^3$ according to the
non-perturbative QCD approaches, for instance, see
Refs.~\cite{dr01,ge68,ga82}, whereas the perturbative vacuum is
chirally symmetric for massless quarks. In the hot and/or dense
medium, the partial restoration of chiral symmetry with reduced
in-medium $<\bar{q}q>$ is expected. The hadronic and electromagnetic
signals for the partial restoration of chiral symmetry have been
suggested in nucleus-nucleus reactions, for instance,
see~\cite{ca99,ko96,fr98}.

Due to the difficulties of the non-perturbative QCD, the in-medium
chiral condensate may be derived alternatively using the QCD sum
rules~\cite{co95},  and/or effective QCD
models~\cite{vo97,ca99,co92,ch01,pe02,ji02}. The Nambu-Jona-Lasinio
models and the linear $\sg$ models are among the well-known examples
of the effective QCD models. Recently, the in-medium chiral
condensate was also exploited in some new
attempts~\cite{mi97,ch07,ki07}. A reduced in-medium chiral condensate
was obtained in these models and approaches. However, an appropriate
description of the nucleonic medium in terms of the quark degree of
freedom is still very difficult. Interestingly, a model-independent
approach to evaluate the in-medium chiral condensate was developed by
applying the Hellmann-Feynman (HF) theorem~\cite{co92}. On the hadron
level, the in-medium chiral condensate can be obtained from the
derivative of the nuclear matter energy density with respect to the
current quark mass. In fact, the in-medium chiral condensate has been
evaluated using the nuclear matter energy density obtained from
various many-body theories. For instance, the Dirac-Brueckner (DB)
approach~\cite{li94,bro96},  the relativistic mean field (RMF)
models~\cite{de95,de96,ma97,ba05},  the quark meson coupling
models~\cite{sa95,wa03}, and   the  effective field
theory~\cite{pl08,ka08}, have all been used in studying the in-medium
chiral condensate. Surprisingly, most of these studies have shown
various hindrances towards the chiral symmetry restoration at high
densities when nucleon-nucleon interactions are taken into account
even in the modified quark-meson coupling model~\cite{ji99}.

In principle, the non-vanishing chiral condensate arises from the
vacuum of the non-perturbative QCD. In particular, the nucleonic
vacuum can be considered as one of hierarchic constructions of the
non-perturbative QCD vacuum. At finite density, the vacuum is
modified by the medium. In dealing with the chiral condensate in
nuclear medium, a proper treatment of the nucleonic vacuum is thus
very important. However, so far very little attention has been paid
to the vacuum corrections in evaluating the in-medium chiral
condensate because usually only effective models are used. In this
work, we explore the in-medium chiral condensate with the vacuum
corrections using the Lagrangian constructed by us
recently~\cite{ji07,jia07} adopting the Brown-Rho (BR) scaling law
for the partial restoration of chiral symmetry in nuclear
matter~\cite{br91,br04}.

The paper is organized as follows. In Section \ref{RHA}, we
renormalize our model with the Brown-Rho scaling in the Hartree
approximation to attain the finite contribution from the Dirac sea.
In section \ref{chiral}, we present the formalism for  the in-medium
chiral condensate using the HF theorem. Results on the Equation of
State (EOS) and in-medium chiral condensate are discussed in Section
\ref{results}. The summary is given in Section \ref{summary}.

\section{Relativistic Hartree model with BR scaling}
\label{RHA}

In the present work, the model Lagrangian with density-dependent
couplings and meson masses is written as~\cite{jia07}
\begin{eqnarray}
{\mathcal L}&=& {\overline\psi}[i\gamma_{\mu}\partial^{\mu}-M
+g^*_{\sigma}\sigma-g^*_{\omega }
\gamma_{\mu}\omega^{\mu}-g^*_{\rho}\gamma_\mu \tau_3 b_0^\mu]\psi
+\frac{1}{2}(\partial_{\mu}\sigma\partial^{%
\mu}\sigma-m_{\sigma}^{*2}\sigma^{2})  \nonumber \\
&& - \frac{1}{4}F_{\mu\nu}F^{\mu\nu}+ \frac{1}{2}m_{\omega}^{*2}\omega_{\mu}%
\omega^{\mu} - \frac{1}{4}B_{\mu\nu} B^{\mu\nu}+
\frac{1}{2}m_{\rho}^{*2} b_{0\mu} b_0^{\mu}- \frac{1}{4}A_{\mu\nu}
A^{\mu\nu}+ {\mathcal L}_{CT}, \label{eq:lag1}
\end{eqnarray}
where ${\mathcal L}_{CT}$ is the  counter term used for
renormalization, $\psi,\sigma,\omega$, and $b_0$ are the fields of
the nucleon, scalar, vector, and isovector-vector mesons, with their
masses $M, m^*_\sg,m^*_\om$, and $m^*_\r$, respectively. The meson
coupling constants and masses with asterisks denote their density
dependence, given by the BR scaling~\cite {ji07,jia07,song01}.  The
density dependence of parameters is described by the scaling
functions that are the ratios of the in-medium parameters to those in
the free space. We take the scaling function for  meson masses as
given in Ref.~\cite{ji07}
\begin{equation}
\Phi(\rho)=1-y\rho/\rho_0,  \label{sc2}
\end{equation}
where the linear dependence on density is similar to the Nambu
scaling~\cite{br04} but with a much smaller coefficient
$y\approx0.1$ suggested by recent experiments~\cite{kek,tap}. Here,
the scaling for the nucleon is not considered since the nucleon mass
already drops in the medium due to the coupling to the $\sg$ meson.
The scaling functions for the coupling constants of scalar and
vector mesons read
\begin{equation}  \label{sc3}
\Phi_\sg(\rho)=\frac{1}{1+x\rho/\rho_0},\hbox{ } \Phi_\r(\rho)=\frac{%
1-y\rho/\rho_0}{1+y_\r\rho/\rho_0}, \hbox{ } \Phi_\om(\rho)=\frac{%
1-y\rho/\rho_0}{1+y_\om\rho/\rho_0},
\end{equation}
which are necessary for an appropriate description of nuclear matter
properties~\cite{ji07}.

The model renormalization is treated in the Hartree approximation.
The Lagrangian for the counter terms reads
\begin{equation}\label{lct}
    {\mathcal L}_{CT}=\af\sg+\frac{1}{2!}\bt\sg^2+\frac{1}{3!}\gm\sg^3+
    \frac{1}{4!}\ld\sg^4,
\end{equation}
where the definition and meaning of the coefficients in ${\mathcal
L}_{CT}$ follow those given in Refs.~\cite{ch77,ch77b} but with the
$g_\sg$ replaced by $g^*_\sg$. These counter terms are used to
renormalize the nucleon scalar self-energy, while no renormalization
procedure is needed for the nucleon vector self-energy due to the
nucleon current conservation. We use the procedure developed by Chin
in Ref.~\cite{ch77b} to renormalize the scalar self-energy and the
energy density. In the following, we just stress the renormalization
for the rearrangement term. The latter is essential for the
thermodynamic consistency in deriving the pressure. In the Hartree
approximation, the expectation value of the rearrangement term
$\Sg^R_0$ is given by
\begin{equation}\label{rre}
\Sg^R_0=<\frac{\pp{\mathcal L}}{\pp \r}>=-\r^2C_\om \frac{\pp
C_\om}{\pp \r}-\r^2\dt^2 C_\r\frac{\pp C_\r}{\pp
    \r}-m_\sg^*\sg^2\frac{\pp m_\sg^*}{\pp \r}+\Sg^C,
\end{equation}
and
\begin{equation}\label{rrec1}
\Sg^C=<\bar{\psi}\psi>\sg\frac{\pp g^*_\sg}{\pp \r}+ \frac{\pp
\af}{\pp \r}\sg+ \frac{1}{2!}\frac{\pp \bt}{\pp
\r}\sg^2+\frac{1}{3!}\frac{\pp \gm}{\pp \r}\sg^3+
    \frac{1}{4!}\frac{\pp \ld}{\pp \r}\sg^4,
\end{equation} with
\begin{equation}\label{ssfe}
 <\bar{\psi}\psi>=-i\sum_{i=p,n}\int\frac{d^4\!k}{(2\pi)^4}{\rm tr} G(k)=
    \r_S-i\sum_{i=p,n}\int\frac{d^4\!k}{(2\pi)^4}{\rm tr} G_F(k).
\end{equation}
Here $C_\om=g_\om^*/m_\om^*$, $C_\r=g_\r^*/m_\r^*$,
$\dt=(\r_p-\r_n)/\r$ is the isospin asymmetry, $\r_S$ is the scalar
density, and $G(k)$ is the nucleon propagator. After the
renormalization, the $\Sg^C$ can be expressed as
\begin{equation}\label{rrec2}
    \Sg^C=\sg\frac{\pp g^*_\sg}{\pp\r}\tilde{\r_S}
    =\sg\frac{\pp g^*_\sg}{\pp\r}\left[\r_S-\sum_{i=p,n}
    \frac{2}{(2\pi)^2}\left({M^*}^3\ln\frac{M^*}{M}+\Sg_SM^2-
    \frac{5}{2}\Sg_S^2M+\frac{11}{6}\Sg_S^3\right)\right],
\end{equation}
where the scalar nucleon self-energy $\Sg_S=M-M^*$ with
$M^*=M-g_\sg^*\sg=M-{g_\sg^*}^2\tilde{\r_S}/{m_\sg^*}^2$. The energy
density and pressure read, respectively,
 \bey  \label{eqe1}
 {\mathcal{E}}&=&\half C_\om^2\r^2+\half C_\r^2 \r^2\dt^2+ \half
 \tilde{C}^2_\sg(M^*-M)^2 +\sum_{i=p,n}
 \frac{2}{(2\pi)^3}\int_{0}^{{k_F}_i}\! d^3\!k~ E^*+\Dt{\mathcal E}\\
 p&=&\half C_\om^2\r^2+\half C_\r^2 \r^2\dt^2- \half
 \tilde{C}^2_\sg(M^*-M)^2 -\Sg^R_0~\r\nnb \\
 & &+
 \frac{1}{3}\sum_{i=p,n}\frac{2}{(2\pi)^3}\int_{0}^{{k_F}_i}\! d^3\!k
 ~\frac{{\bf k}^2}{E^*}-\Dt{\mathcal E},
\label{eqp1}
 \eey
where $\tilde{C}_\sg=m_\sg^*/g_\sg^*$, and $\Dt{\mathcal E}$ is the
finite vacuum correction~\cite{ch77,ch77b}
\begin{equation}\label{evac}
\Dt{\mathcal E}=-\frac{1}{4\pi^2}\left( M^{*4}\ln \frac{M^*}{M}
+M^3\Sg_S-\frac{7}{2}M^2\Sg_S^2+\frac{13}{3}M\Sg_S^3-
\frac{25}{12}\Sg_S^4\right).
\end{equation}

\section{In-medium chiral condensate}
\label{chiral}

The HF theorem transmits a parameter dependence of the Hamiltonian to
that of the system energy. For the QCD Hamiltonian, one may divide it
into a chirally invariant part and a part depending on the current
quark mass, namely ${\mathcal H}_{QCD}={\mathcal H}_0+m_q \bar{q}q$.
By taking $m_q$ as the very parameter, the HF theorem reads
\begin{equation}\label{hf}
    <\psi|\frac{\pp H_{QCD}}{\pp m_q}|\psi>=
    \frac{\pp}{\pp m_q}<\psi|H_{QCD}|\psi>.
\end{equation}
Since the vacuum expectation value (VEV) $<0| H_{QCD}|0>$ is
subtracted to obtain the total energy of nuclear matter, the
in-medium chiral condensate $<\bar{q}q>_\r$  is given as~\cite{co92}
\begin{equation}\label{cc1}
    <\bar{q}q>_\r=<\bar{q}q>_0+\frac{\pp {\mathcal E}}{\pp m_q}.
\end{equation}
Here, ${\mathcal E}$ is given in Eq.(\ref{eqe1}), and can be written
more generally as
\begin{equation}\label{eqe2}
    {\mathcal E}=(M+E/A)\r,
\end{equation}
where $E/A$ is the binding energy per nucleon. Both hadron masses and
meson-nucleon coupling constants depend on the current quark mass.
However, based on an analysis in the linear $\sg$ model, the
derivative of the coupling constant with respect to the current quark
mass is smaller than that of the hadron masses by a factor
$\sg_N/M$~\cite{bro96}, with the pion-nucleon sigma term
$\sg_N\approx45~{\rm MeV}$. Thus, we neglect the dependence of the
coupling constants on the current quark mass. The in-medium chiral
condensate is then given by
\begin{eqnarray}\label{cc1}
    <\bar{q}q>_\r&=&<\bar{q}q>_0+\left(~\frac{d M}{d m_q} +
    \frac{\pp (E/A)}{\pp M}\frac{dM}{d m_q}
    \sum_{\sg,\om,\r}\frac{\pp (E/A)}{\pp m_i^*}\frac{dm_i^*}{d
    m_q}\right)\r.
\end{eqnarray}
The current quark mass derivative of the nucleon mass is expressed in
terms of the $\sg_N$ as~\cite{co92,li94}
\begin{equation}\label{sgm}
\frac{d M}{d m_q}=\frac{\sg_N}{m_q}.
\end{equation}
For the vector ($\om$ and $\r$) mesons, one may resort to the
constituent quark model and give the vector meson sigma term using
the following relation
\begin{equation}\label{sgv}
\frac{m_q}{m_V^*}\frac{d m^*_V}{d m_q}=\frac{m_q}{M}\frac{d M}{d
m_q}=\frac{\sg_N}{M}.
\end{equation}
This approximation has been widely used in the
literatures~\cite{co92,li94,bro96,ji99}. The determination from the $\pi N$
scattering  gives $\sg_N=45\pm 7$ MeV~\cite{ga91}. This central value can be
reached in the constituent quark model when the quark confinement condition is
imposed~\cite{pe02}. Recently, in Ref.~\cite{ka08} the two-pion exchange
processes including the intermediate $\Delta$ isobar excitations instead of the
surrogated $\sg$-meson exchange were considered, and a hindered tendency toward
the chiral symmetry restoration at high densities was still shown. In these
chiral effective field theories~\cite{pl08,ka08}, the short-range
Nucleon-Nucleon (NN) interactions only play a minor role on the change of the
in-medium quark condensate, while the long- and intermediate-range NN
interactions from one- and two-pion exchanges dominate the hindrance. This
observation is instructive, though different from that in the
one-boson-exchange approaches~\cite{li94,bro96,de95,de96,ma97,ba05,sa95,ji99}.
However, the strict renormalization for the low-energy physics and contact
interactions in these chiral effective field theories is nevertheless
imperative to further judge whether the additional degrees of freedom are
necessary for describing the short-range interactions. Though the progress of
renormalization has been made recently~\cite{no05,va06,en08}, this is still
quite challenging. In these chiral effective theories that were built upon the
chiral pertubative theory, the vacuum is alienated in two folds: First, the
pion as the Goldstone boson that resides in the coset space is not accompanied
by other boson degrees of freedom that carry the chirally invariant vacuum in
the conserved subgroup space. Second, in a non-relativistic treatment the
nucleonic vacuum is out of touch. In fact, suitable renormalization scheme of
the baryonic vacuum allows one to include consistently the vector mesons as
explicit degrees of freedom in the Lorentz-invariant baryon chiral perturbative
theory~\cite{fu03}.  Since the chiral symmetry restoration is expected to occur
at the short-range region with a chirally invariant vacuum, the present
approach is of close relevance to dealing with the in-medium chiral condensate
regarding the importance of the vacuum correction and the important role of
short-range $\om$-meson exchange. As we focus on the vacuum correction, the
relation (\ref{sgv}), though phenomenological, can provide appropriate and
referential information.

The situation with the scalar $\sg$ meson is more involved since it
has a broad mass spectrum starting roughly above $2m_\pi$, as a
result of the equivalence  of a variety of two-pion exchanges. As
known in the linear $\sg$ model of Gell-Mann and Levy~\cite{ge60},
the $\sg$ meson is regarded as the chiral partner of the Goldstone
boson ($\pi$ meson). In this model the chiral symmetry and its
spontaneous breaking are well manifested. Although one can have some
other choices to scale the sigma term for the $\sg$
meson~\cite{co92,li94,bro96,ji99}, we stress here that the symmetry
of chirality is important and thus in our model that is designed to
respect the chiral symmetry restoration at high densities  we persist
in using the sigma term for the $\sg$ meson obtained from the linear
$\sg$ model, and at tree level it is~\cite{ch01,bro96,bir93}
\begin{equation}\label{sgsg}
\frac{m_q}{m_\sg^*}\frac{d m^*_\sg}{d
m_q}=\frac{3}{2}\frac{\sg_N}{M}.
\end{equation}
The use of this relation is consistent with present relativistic
Hartree approach which is also at tree level, though there can be
high-order corrections to this relation. Here, the $\sg$-meson mass
is taken to be 590 MeV. It was obtained by fitting ground-state
properties of many nuclei with the same scaling functions used in our
previous work~\cite{jia07}. Using relations (\ref{sc2}), (\ref{sgm}),
(\ref{sgv}), (\ref{sgsg}) and the Gell-Mann-Oakes-Renner relation
$m_\pi^2 f^2_\pi=-m_q<\bar{q}q>_0$~\cite{ge68}, the in-medium chiral
condensate is eventually given by
\begin{eqnarray} \label{cc2}
\frac{<\bar{q}q>_\r}{<\bar{q}q>_0}&=&1-\frac{\sg_N}{m_\pi^2
f^2_\pi}\r\left(1+\frac{\pp (E/A)}{\pp M} + \frac{\pp (E/A)}{\pp
m_\sg^*}\frac{3m_\sg}{2M}\Phi(\r)
-C_\om^2\frac{\r}{M}-C_\r^2\frac{\r\dt^2}{M}\right),
\end{eqnarray}
with the $\pi$-meson decay parameter $f_\pi=93$ MeV and the
$\pi$-meson mass $m_\pi=138$ MeV. The first term in the bracket is
the leading-order model-independent contribution. In this work, we
focus on the chiral condensate in symmetric nuclear matter where the
$\r$ meson does not contribute. We also do not consider the $\dt$,
$\pi$ and $\eta$ mesons, as they do not contribute to the symmetric
nuclear matter either in the Hartree approximation. In fact, within
the Dirac-Brueckner approach, their contributions to the in-medium
condensate are about two orders of magnitude smaller than those of
$\sg$ and $\om$ mesons~\cite{li94,bro96}. Note that the role of the
$\pi$ meson in these relativistic one-boson-exchange theories is
different from that in the effective field theories~\cite{pl08,ka08},
where the perturbative expansion with the unique $\pi$ meson degrees
of freedom results in the multi-pion exchanges~\cite{no05}.

\section{Results and discussions}
\label{results}

Before presenting our results on the in-medium chiral condensate, we
discuss more about the determinations of model parameters. In our
previous work~\cite{ji07,jia07}, besides the saturation properties of
nuclear matter, namely, the incompressibility $\kappa=230$ MeV at
saturation density $\r_0=0.16~{\rm fm}^{-3}$, we also use as a
constraint the EOS of high density nuclear matter determined from
analyzing the collective flow of relativistic heavy-ion
collisions\cite{da02}. Here, we use the same scaling functions of
meson masses and meson-nucleon coupling constants as for the model
SL1 constructed in Refs.~\cite{ji07,jia07}. The model SL1, however,
does not have the vacuum renormalization. The resulting new parameter
set RSL1 and the corresponding saturation properties are tabulated in
Table \ref{t:t1}. Compared to the parameter set SL1, in the RSL1 the
coupling constant of the $\om$ meson is significantly reduced due to
the vacuum contribution. Consistent with the finding in
Ref.~\cite{ch77b}, the nucleon effective mass at saturation density
increases significantly due to the vacuum correction. This means that
the in-medium nucleon feels a scalar repulsion from the Dirac sea,
and it prevents nucleons in the Fermi sea from falling to the
fully-occupied Dirac sea by sustaining a larger mass gap.
\begin{table}[htb]
\caption{Parameter sets RSL1, RSL2 and SL1.  The vacuum hadron masses
are $M=938$MeV, $m_\sg=590$MeV, $m_\om=783$MeV and $m_\r=770$MeV. The
coupling constants given here are those at zero density.  The
symmetry energy is fitted to 31.6MeV at saturation density.
\label{t:t1}}
 \begin{center}
    \begin{tabular}{ c c c c c c c c c c c c c}
\hline\hline &$g_\sg$ & $g_\om$ & $g_\r$ & $y$ & $x$ &$y_\om$&
$y_\r$&$\kappa$(MeV)   & $M^*/M$     \\
\hline
RSL1  &8.7376 &7.8476 &4.0371 &0.126 & 0.2831 & -&-&230 &  0.788    \\
RSL2  &8.3034 &6.9439 &3.9878 &0.126 & 0.2988 &-0.0225&- &230 &  0.810 \\
 SL1  &10.1937&10.4634&3.7875 &0.126 & 0.234 &-&- &230 &  0.679 \\
\hline\hline
\end{tabular}
\end{center}
\end{table}

\begin{figure}[tbh]
\begin{center}
\vspace*{-20mm}\includegraphics[width=0.7\textwidth]{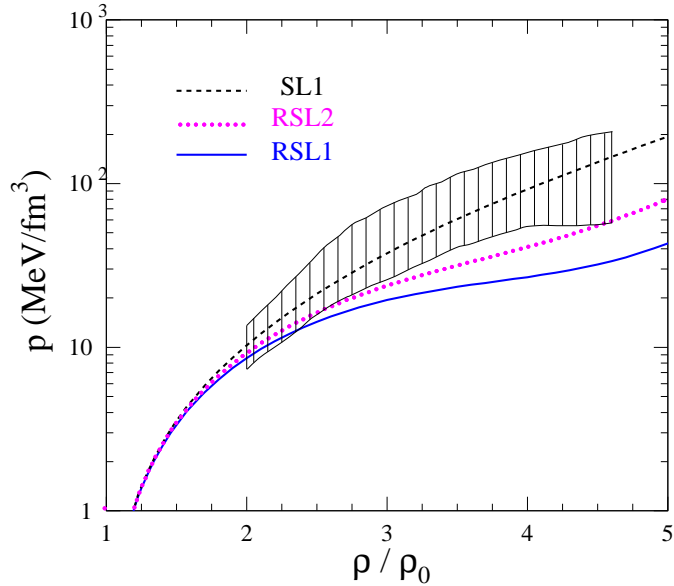}
\vspace*{-10mm} \end{center} \caption{(Color online) The pressure for
different models versus density in symmetric nuclear matter. The
shaded region is given by experimental error bars\cite{da02}.
\label{f:f1}}
\end{figure}
Shown in Fig.~\ref{f:f1} are the EOSs of symmetric nuclear matter for the RMF
model SL1 and the relativistic Hartree models. The shaded area indicates the
experimental constraint on the high-density EOS\cite{da02}. The model
parametrization in the SL1 was selected such that the resulting EOS passes
through the middle of the constraint. Comparing to the EOS with the SL1, it is
seen that the pressure with the RSL1 is much reduced at high densities due to
the vacuum correction. In order to bring the predicted pressure closer to the
experimental region, we use another parameter set RSL2 that is adjusted to have
a larger nucleon effective mass at saturation density, as listed in Table
\ref{t:t1}. As shown in Fig.~\ref{f:f1}, the moderate readjustment of
parameters results in a significant increase of pressure at high densities
where the rearrangement term plays an important role.

We have noticed that the energy density at high densities is
dominated by the repulsion provided overwhelmingly by the $\om$
meson. This is the main origin of the hindrance toward a chiral
symmetry restoration observed using energy densities predicted by
most meson-exchange many-body  theories. In the simple RMF models,
the repulsion increases linearly with density. After taking into
account the higher-order correlations beyond the mean-field
approximation within the DB approach, though the EOS at high
densities was softened, the increasing tendency of the in-medium
chiral condensate was not overturned~\cite{li94,bro96}. On the other
hand, this may also be understandable  since these approaches are not
really expected to give a correct description of chiral condensate at
least near the critical density for the chiral symmetry restoration.
It was found very recently within the Brueckner approach with chiral
limit that nucleon-nucleon interactions lead to a weaker reduction of
the in-medium chiral condensate in the density region where the
effective field theory  is applicable~\cite{pl08}, while little
attention was paid to the saturation properties. The decreasing
tendency of the in-medium chiral condensate can be obtained in RMF
models by including the density-dependent coupling constants, namely
the density-dependent ratios $\tilde{C}_\sg$ and $C_\om$~\cite{de95}.
In this work, instead of modifying the ratios $\tilde{C}_\sg$ and
$C_\om$, we will investigate the role played by the vacuum
corrections.

\begin{figure}[tbh]
\begin{center}
\vspace*{-20mm}\includegraphics[width=0.7\textwidth]{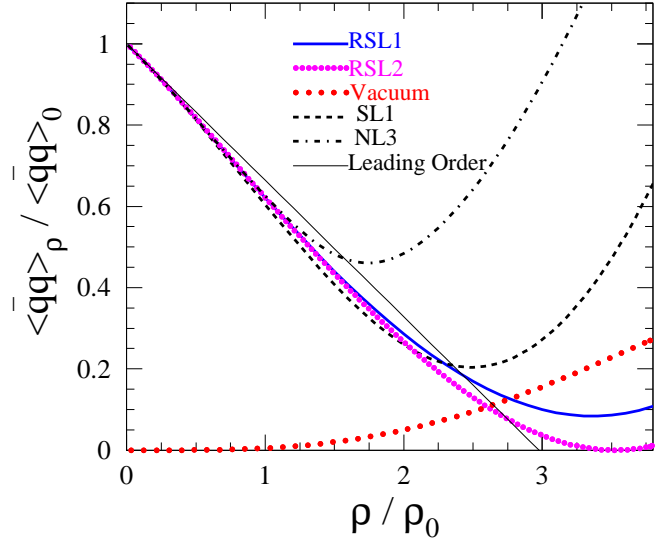}
\vspace*{-10mm} \end{center} \caption{(Color online) In-medium chiral
condensate for various models. The "vacuum" is calculated with the
RSL2. \label{f:f2}}
\end{figure}
In Fig.~\ref{f:f2}, we show the in-medium chiral condensate
evaluated using Eq.(\ref{cc2}) with the RMF models NL3~\cite{nl3}
and SL1 and the relativistic Hartree models  RSL1 and RSL2. Compared
to the NL3 result, the in-medium chiral condensate with the SL1 is
largely reduced at higher densities. This can be attributed to a
much softened EOS with the SL1~\cite{ji07}, compared to that with
the NL3~\cite{da02}. At high densities, however, the chiral
condensate with the SL1 increases with density given the fact that
the constant ratio $C_\om$ in SL1 does not suppress the vector part
of the energy density. The latter is quadratic in density, also see
Eqs.(\ref{eqe1}) and (\ref{cc2}). It is seen that the mass dropping
given by the BR scaling is not sufficient to reduce the chiral
condensate to zero due to the nucleon-nucleon interactions in
nuclear matter.

As the chiral condensate arises from the vacuum, the appropriate
description of the Dirac vacuum contribution is imperative. The
curve in the figure denoted by "vacuum" is calculated from the
vacuum term $\Dt{\mathcal E}$ in Eq.(\ref{eqe1}). It is seen that
the vacuum contribution to the in-medium chiral condensate becomes
increasingly more important at higher densities. Moreover, the
vacuum correction in the energy density also includes the part that
modifies the nucleon effective mass. In the renormalized model RSL1
the chiral condensate further goes down with the increasing density.
We note that the vacuum energy density $\Dt{\mathcal E}$ is positive
and the scalar field $\sg$ that provides a big attraction is reduced
due to the vacuum correction. Thus, the vacuum correction actually
gives rise to a scalar repulsion that impairs the role of the vector
repulsion provided by the $\om$ meson. As a result, the much reduced
$g_\om$ with a consistent but smaller reduction of $g_\sg$, subject
to the given saturation properties, results in a softened EOS and
brings down the chiral condensate at high densities in the RSL1,
compared to that with the SL1.

Interestingly, when we use the parameter set RSL2 that is adjusted to
have an EOS more close to the experimental constraint at high
densities, see Fig.~\ref{f:f1}, the in-medium chiral condensate
decreases further to zero by coincidence. As shown in
Eqs.(\ref{eqe1}) and (\ref{eqp1}), the vacuum correction
$\Dt{\mathcal E}$ contributes oppositely to the energy density and
the pressure. This provides the intrinsic mechanism for the different
changing tendencies in the energy density and the pressure,
intriguing the shift going from the RSL1 to the RSL2 as observed in
Figs.~\ref{f:f1} and \ref{f:f2}. Numerically, we find that a larger
cancellation of contributions from the scalar and vector mesons in
the RSL2, compared to the RSL1, is responsible for the big half of
the reduction of the chiral condensate from the RSL1 to the RSL2,
whereas the rest reduction arises directly from the variation of the
vacuum correction. This further shows a direct link between the
in-medium chiral condensate and the vacuum correction from the Dirac
sea. With the vacuum correction, the nucleon effective mass drops
more slowly in nuclear matter. As it was also stressed in the
modified RMF models~\cite{de95} and the Brueckner approach with the
chiral limit~\cite{pl08}, there is a decoupling between the
reductions of the nucleon effective mass and the chiral condensate at
high densities. Actually, this decoupling exists in most hadronic
approaches where the in-medium chiral condensate increases whereas
the nucleon effective mass drops at high densities.

In this work, we have concentrated on investigating effects of the
vacuum corrections on the in-medium chiral condensate. However, if
the chiral condensate does not drop sufficiently at high densities,
other corrections, such as, high-order correlation contributions, may
help further bring it down. Another simple way is to reduce the
in-medium vector coupling constant. The expectation of vanishing
chiral condensate at high densities would actually add some
constraints on the in-medium scaling of the vector coupling constant
in the framework of the BR scaling. It is also interesting to point
out that the parameter used in the scaling function for the meson
masses is consistent with that extracted from recent low-density
experimental data, see Ref.~\cite{ji07}. However, it may be risky to
apply the currently used scaling function having a linear density
dependence of meson masses to high densities. In Ref.~\cite{br04},
different scalings were suggested for the regions below and above the
saturation density. Moreover, one may consider the use of  the
nucleon mass scaling. Theoretical uncertainty may also exist in the
approximations to determine the sigma terms for the vector and scalar
mesons. Certainly, all these details deserve further investigations.
Nevertheless, regardless of the very details and the uncertainties
that ebb and flow, a definite fact is that the vacuum correction
brings out a new mechanism favoring the reduction of the in-medium
chiral condensate.

\section{Summary}
\label{summary}

In summary, we have studied the vacuum corrections in the
relativistic Hartree model using the density-dependent
parameterizations according to the BR scaling. The in-medium chiral
condensate have been investigated using the Hellmann-Feynman theorem
together with phenomenological relations for the current quark mass
derivative of hadron masses. Constrained by the saturation
properties, the vector repulsion is reduced by the scalar repulsion
arising from the Dirac sea. This results in the softened EOS and the
reduction of the in-medium chiral condensate at high densities. Also,
the vacuum correction from the Dirac sea plays  a direct and
important role in generating the in-medium chiral condensate. The
interplay between the vacuum corrections to the energy density and
the pressure provides a mechanism favoring the chiral symmetry
restoration in dense matter.

\section*{Acknowledgement}
The work was supported in part by the US National Science Foundation
under Grant No. PHY-0652548,  PHY-0757839   and  the Research
Corporation  under the Award No. 7123. One of authors WZJ also thanks
the partial support from the NNSF of China under Grant Nos. 10405031,
10575071 and 10675082, the KIP of the Chinese Academy of Sciences
under Grant No. KJXC3-SYW-N2, and the China MSBRDP under Contract No.
2007CB815004.

\end{document}